# Inverse-designed Photonic Computing Core for Parallel Matrix-vector Multiplication

KAIYUAN WANG[1,2], YUNLONG LI[1,2], TIANGE WU[1,2], DEMING LIU[1,2], SHUANG ZHENG[1,2,*], MINMING ZHANG[1,2,3,*]

[1]School of Optical and Electronic Information, Huazhong University of Science and Technology, Wuhan, Hubei 430074, China
[2]National Engineering Research Center for Next Generation Internet Access System, Wuhan, Hubei 430074, China
[3]Wuhan National Laboratory for Optoelectronics, Wuhan, Hubei 430074, China
* zshust@hust.edu.cn and mmz@hust.edu.cn

**Abstract:** On-chip optical neural networks (ONNs) have recently emerged as an attractive hardware accelerator for deep learning applications, characterized by high computing density, low latency, and compact size. As these networks rely heavily on massive matrix multiplication, photonic computing cores for matrix computation become crucial components for on-chip ONNs, which harness the degree of freedoms (DOFs) in photonics including space, wavelength and mode dimensions. However, previous photonic computing devices have not fully utilized the orthogonality and the conversion characteristic of the waveguide modes, which as we show here, allows for the simultaneous parallel computing of several independent matrix-vector multiplications within the same device. In this work, we propose an inverse-designed photonic computing core for parallel matrix-vector multiplication. The matrices are implemented through a mode conversion process, where the input fundamental modes are simultaneously converted into several orthogonal output modes. Specifically, we target the complex-valued conversion matrices between input and output modes and inversely design the dielectric distribution within the device to achieve parallel matrix-vector multiplication. As a demonstration, the proposed photonic computing core supports simultaneous parallel computing of two independent matrix-vector multiplications, with an ultra-compact footprint (4.8×2.88 µm²) and high computing precision (relative error < 8%) at 1550 nm wavelength. The inverse-designed photonic computing devices hold great potential for high-performance on-chip ONNs with low energy consumption and high computing density.

## 1. introduction

Artificial neural networks (ANNs) have shown remarkable capabilities in the applications of advanced technology such as autonomous driving, natural language processing, and medical diagnostics [1]. Optical neural networks (ONNs) have emerged as a promising technology for the implementation of ANNs, since optical devices exhibit inherent high computing density, low latency, and compact size [2]. As the computing process of ONNs heavily relies on massive matrix-vector multiplication during both the feedforward and backpropagation phases, photonic cores for matrix multiplication have become critical for ONNs [3]. Specifically, a photonic core for matrix multiplication with high-performance parallelism can greatly improve the computing density and decrease the energy consumption of ONNs.

The photonic core harnesses the degree of freedoms (DOFs) in photonics including space, wavelength, and mode to achieve matrix-vector multiplication. Firstly, two DOFs are utilized, that is, space and wavelength. One such scheme involves arrays of cascaded Mach-Zehnder interferometers (MZIs) that utilize coherent light for data transmission, with thermo-optic phase shifters serving as dynamic weighting elements [4] . Another scheme utilizes various wavelengths of light as distinct data channels, employing tunable microring resonators (MRRs) for the modulation of these channels, which is compatible with wavelength-division multiplexing (WDM) [5] . Recently, a photonic core based on photonic crystal nanobeam cavities (PCNCs) has been realized, which achieves a remarkably compact size and demonstrates a significant increase in thermal tuning efficiency compared to microring resonators [6] . Secondly, the DOF of waveguide mode is further utilized. Based on mode division multiplexing (MDM), one scheme employed a mode multiplexer, a multimode splitter, mode demultiplexers, modulators, and combining elements, which realize a photonics core for 4×4 matrix multiplication [7] . Another scheme is based on WDM-compatible MDM. This work presents real-number-field computations for both positive and negative domains and demonstrates a computing density of 1.37 TOPS/mm$^2$ under the 22.38 Gbaud modulation rate for image convolution and pattern detection. The characteristics of MDM reduce the use of lasers at different wavelengths, thereby reducing complexity and cost for integration [8] . Though the DOF of waveguide mode has shown great potential, previous photonic cores are unable to support simultaneous parallel computing of several independent matrix multiplications within the same device, which significantly restricts the computing density and the energy consumption.

In this work, we propose a photonic core for parallel matrix multiplications by fully utilizing the orthogonality and the conversion characteristic of the waveguide modes, where the input fundamental modes are simultaneously converted into several orthogonal output modes. Specifically, we employ the photonic-crystal-like (PhC-like) subwavelength structure and applied the adjoint method to inversely design the dielectric distribution, which has shown powerful capacity in flexible control of waveguide modes [9-11]. To achieve parallel multiplication of several independent matrices, we target the complex-valued conversion matrices between input and output modes. In this design target, the complex amplitude transmission between the input and output modes corresponds to the complex elements in the conversion matrices. As a demonstration, the proposed photonic computing core supports simultaneous parallel computing of two independent matrix-vector multiplications, with an ultra-compact footprint (4.8×2.88 µm²) and high computing precisions (relative error < 8%) at 1550 nm wavelength. The inverse-designed photonic computing devices hold great potential for high-performance on-chip ONNs with low energy consumption and high computing density.

## 2. Design Principle

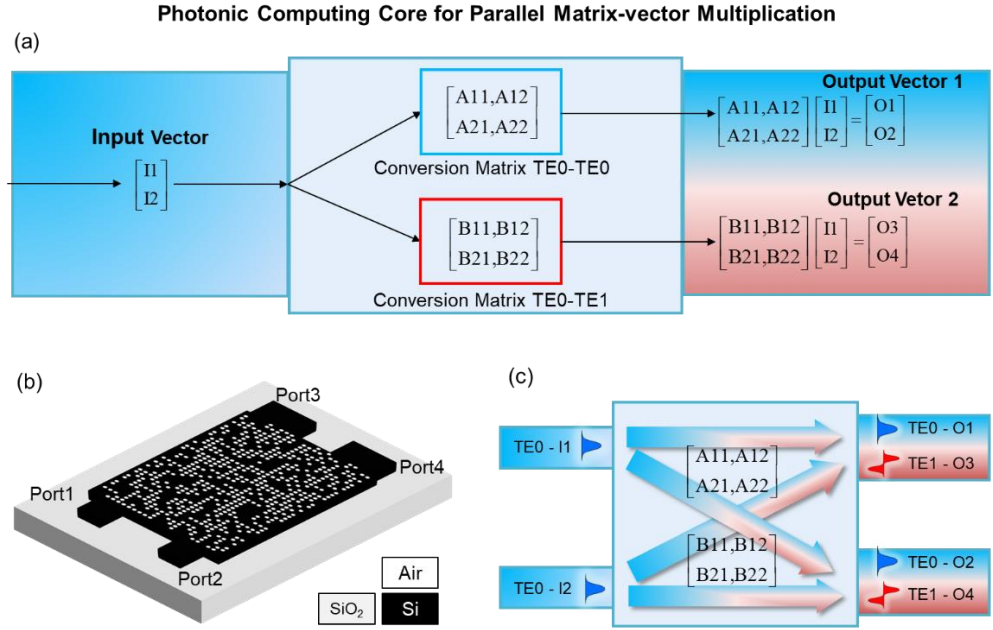

Figure 1: (a) Schematic of the capability of our photonic core for parallel matrix multiplication, enabling simultaneous parallel computing of two independent matrix-vector multiplications. (b) 3D schematic of the photonic core structure, constructed with the photonic-crystal-like (PhC-like) subwavelength structure, featuring two input ports, two output ports, and an inversely designed region. (c) Schematic of the implementation of our photonic core, indicating that the matrix-vector multiplications correspond to the mode conversion between the input modes and the output modes.

Figure 1(a) illustrates our photonic core's capability to perform parallel computing of two independent matrix-vector multiplications. The same input vector is simultaneously multiplied by two independent conversion matrices to yield two distinct vectors. Specifically, the formulas for the processes are as follows:

$$\begin{bmatrix} O1 \\ O2 \end{bmatrix} = \begin{bmatrix} A11, A12 \\ A21, A22 \end{bmatrix} \begin{bmatrix} I1 \\ I2 \end{bmatrix}, \begin{bmatrix} O3 \\ O4 \end{bmatrix} = \begin{bmatrix} B11, B12 \\ B21, B22 \end{bmatrix} \begin{bmatrix} I1 \\ I2 \end{bmatrix} \quad (1)$$

Within these formulas, I1 and I2 represent the complex elements in the input vector, while O1,O2 and O3,O4 represent the complex elements in the output vector 0 and the output vector 1 respectively. Aij and Bij (where i, j ∈ {1, 2}) represent the complex elements in the conversion matrix TE0-TE0 and the conversion matrix TE0-TE1 respectively. In the parallel matrix-vector multiplication process, output vector 0 is generated by multiplying the input vector by the TE0-TE0 conversion matrix. Simultaneously, output vector 1 is generated by multiplying the input vector by the TE0-TE1 conversion matrix. As depicted in Fig. 1(b), our photonic core is constructed on a photonic-crystal-like (PhC-like) subwavelength structure. Specifically, our photonic core is composed of two single-mode input waveguides with a width of 450 nm, and two output waveguides supporting two modes with a width of 900 nm. The inverse design region of 4.8×2.88 µm² is discretized into 24 × 40 pixels with a central hole of 90 nm diameter and 220 nm depth.

Figure 1(b) illustrates the implementation of our photonic core, indicating that the matrix-vector multiplications correspond to the mode conversion between the input modes and the output modes. Specifically, the parallel matrix-vector multiplication is implemented through a mode conversion process, where the input fundamental modes are simultaneously converted into several orthogonal output modes. In the input waveguides, the fundamental modes TE0 - I1 and TE0 - I2 carry the input vector. In the output waveguides, the fundamental modes TE0 - O1 and TE0 - O2 carry the output vector 0, while first-order modes TE1 – O3 and TE1 – O4 carry the output vector 1. In our design principle, the complex amplitude transmission between the input and output modes corresponds to the complex elements in the conversion matrices. By using the digitized adjoint method, we can target the complex-valued conversion matrices and inversely design the dielectric distribution in our device to achieve parallel matrix-vector multiplication.

## 3. Photonic Core for Parallel Multiplication of Diagonal and Zero Matrices

Here, we demonstrate our optimization process with a design target for the parallel multiplication of diagonal and zero matrices. Specifically, the formulas for the design target are as follows:

$$\begin{bmatrix} O1 \\ O2 \end{bmatrix} = \begin{bmatrix} -0.5+0i, 0+0i \\ 0+0i, -0.5+0i \end{bmatrix} \begin{bmatrix} I1 \\ I2 \end{bmatrix}, \begin{bmatrix} O3 \\ O4 \end{bmatrix} = \begin{bmatrix} 0+0i, 0+0i \\ 0+0i, 0+0i \end{bmatrix} \begin{bmatrix} I1 \\ I2 \end{bmatrix} \quad (2)$$

To fulfill the design target, we define the figure of merit[] as follows: FOM =α·Fperf+ β·Fbina. The first term Fperf = γ·{[A11 - (0 + 0i) ] + [A12 - (0 + 0i)] + [A21 - (0 + 0i)] + [A22 - (0 + 0i)]} + (1-γ)·{[B11 - (-0.5 + 0i)] + [B12 - (0 + 0i)] + [B21 - (0 + 0i)] + [B22 - (-0.5 + 0i)]} is the objective function of the device performance, which can force the complex conversion matrix Aij, Bij to zero matrix and diagonal matrix respectively. The second term

Fbina =$\{1- \Sigma(\varepsilon_i – \varepsilon_{air})(\varepsilon_i – \varepsilon_{si})\}$ is the objective function of the binarization, which can force the relative permittivities distribution of holes to a binary distribution. The $\varepsilon_i$, $\varepsilon_{air}$, and $\varepsilon_{si}$ represent the dielectric constants of pixels, air and silicon respectively. α, β, γ are the hyper-parameters which control the weight of terms in the figure of merit. Here, we set α, β, and γ as 0.5, 0.5, and 0.5 respectively.

We can execute our optimization process by calculating the gradient of the figure of merit (dFOM/d$\varepsilon_i$) to optimize the dielectric constants within the pixels. Specifically, we initialize the dielectric constants of all pixels to an intermediate value between silicon and air. Then the dielectric constants distribution within each pixel is iteratively updated by $\varepsilon_i^{new} = \varepsilon_i^{old} + \Delta \cdot dFOM/d\varepsilon_i^{old}$, where the $\varepsilon_i^{new}$, $\varepsilon_i^{old}$ represent the dielectric constant after update and before update. Δ is a hyper-parameter that controls the convergence speed and we set Δ as 0.05. During the iterations, dielectric constants may become smaller than air or larger than silicon; they are subsequently clipped to the values corresponding to air or silicon, respectively. The evolution of our device pattern is shown in Fig. 2(a). As the optimization iterations progressed, we observed a gradual modification in the pixel distribution of the pattern at iterations 50 and 100, with the dielectric constants converging toward silicon or air. We stopped the optimization iterations at iteration 200, where the distribution of dielectric constants closely approximates silicon or air. Then we fully binarize this pattern to obtain the final pattern, using the midpoint between the dielectric constants of silicon and air as the threshold. Specifically, we set pixels with dielectric constants above this threshold to silicon and those below to air, thus obtaining the final pattern in which the dielectric constants of each pixel are either silicon or air.

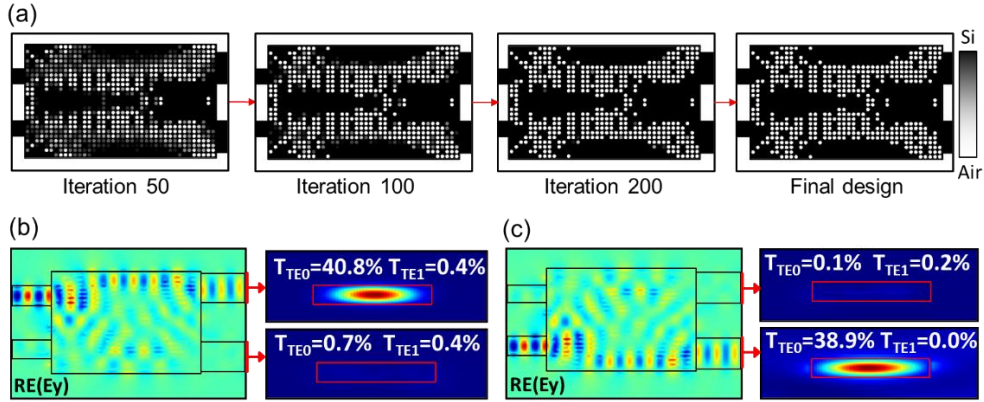

Figure 2: (a) Evolution of the dielectric constants in the pattern during the iterative optimizing process. (b) Simulation results of the final pattern which include the electric fields, power distributions, and normalized transmissions.

To demonstrate the mode conversion characteristics of our device, we conducted finite-difference time-domain (FDTD) simulations of the electric field in the final pattern, the normalized power distributions of the output waveguides, and the normalized transmissions of TE0 and TE1 modes. As Figure 2(b) shows, the TE0 mode is injected into the input waveguide (upper port or lower port) and mode conversions occur in the inverse-designed region. In output waveguide, the normalized transmission of TE1 mode is about zero and the the normalized transmission of TE0 mode is about forty percent that of the input. The simulation results indicate that the mode conversions between input modes and output modes are consistent with our design principle.

To further evaluate the computing precisions of our device, we conducted FDTD simulations to obtain simulation values of each element in matrices. Figure 3(a) illustrates a polar plot for the comparison between design targets and simulation values. Elements within the matrices are differentiated by colors, with targets represented by circles and simulations by crosses. The plot reveals that simulation values closely align with the design targets. In Figure

3(b), the image consists of three sections demonstrating the design target, simulation values, and their relative error within the conversion matrix TE0-TE0. The simulation values closely approximate the design targets with minor deviations. The rightmost section is a bar graph displaying the relative errors for the elements, which are below 2.2%. Similarly, Figure 3(c) displays the conversion matrix TE0-TE1. The simulation values exhibit slight deviations compared to the design targets. The relative error for the four elements ranges from 1.0% to the highest at 4.4%. Overall, the graphical information indicates that the simulation closely matches the design targets with small relative errors, highlighting the high computing precision of our photonics core.

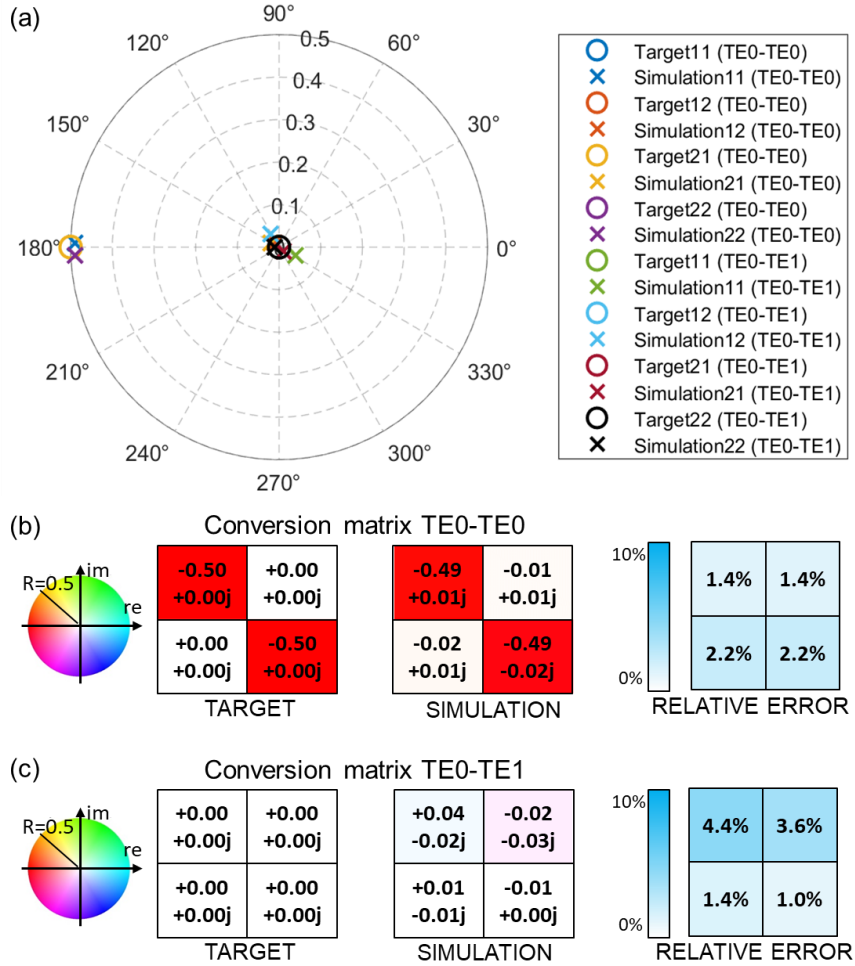

Figure 3: (a) Polar plot representation of the design targets and simulation values for TE0-TE0 and TE0-TE1 mode conversions matrix. (b) Statistical diagram and relative error analysis diagram for TE0-TE0 conversion matrix. (c) Statistical diagram and relative error analysis diagram for TE0-TE1 conversion matrix.

## 4. Photonic Cores for Parallel Multiplication of Two Arbitrary Matrices

In the previous section, we introduced our design principle and demonstrated a photonic core capable of parallel multiplication for diagonal and zero matrices. To validate the scalability of our design principle, we further designed photonic cores capable of performing parallel multiplication on two arbitrary matrices. This confirms the scalability of our design principle for performing parallel computing with any specified matrix. Specifically, we target complex-valued conversion matrices at random to demonstrate parallel computation for arbitrary matrices, showing two design cases here. The design targets for the two cases are formulated as follows:

$$\begin{bmatrix} O1 \\ O2 \end{bmatrix} = \begin{bmatrix} 0.1+0.15i, 0.15-0.1i \\ 0.15+0i, 0+0.25i \end{bmatrix} \begin{bmatrix} I1 \\ I2 \end{bmatrix}, \begin{bmatrix} O3 \\ O4 \end{bmatrix} = \begin{bmatrix} -0.2+0i, 0.2-0.15i \\ 0+0.15i, 0.1+0.2i \end{bmatrix} \begin{bmatrix} I1 \\ I2 \end{bmatrix} \quad (3)$$

$$\begin{bmatrix} O1 \\ O2 \end{bmatrix} = \begin{bmatrix} 0.25+0.2i, -0.2-0.15i \\ 0-0.35i, 0.25+0.15i \end{bmatrix} \begin{bmatrix} I1 \\ I2 \end{bmatrix}, \begin{bmatrix} O3 \\ O4 \end{bmatrix} = \begin{bmatrix} -0.2+0i, 0-0.15i \\ 0+0i, 0.2+0.2i \end{bmatrix} \begin{bmatrix} I1 \\ I2 \end{bmatrix} \quad (4)$$

Formula 3 and Formula 4 respectively represent the design target for case 1 and case 2. In case 1, we define the figure of merit as follows: FOM =0.5·Fperf+ 0.5·Fbina. In the first term, Fperf =0.5 {[A11 - (0.10 + 0.15i) ] + [A12 - (0.15 – 0.10i)] + [A21 - (0.15 + 0.00i)] + [A22 - (0.00 + 0.25i)]} – 0.5 {B11 - (-0.20 + 0.00i)] + [B12 - (0.20 – 0.15i)] + [B21 - (0.00 + 0.15i)] + [B22 - (0.1 + 0.2i)]}. The second term is set as Fbina = 1- Σ($\varepsilon_i$ – $\varepsilon_{air}$) ($\varepsilon_i$– $\varepsilon_{si}$). In the case 2 , we define the figure of merit as follows: FOM =0.5·Fperf+ 0.5·Fbina. The first term Fperf = 0.5 {[A11 - (0.25 + 0.20i) ] + [A12 - (-0.20 – 0.15i)] + [A21 - (0.00 - 0.35i)] + [A22 - (0.25 + 0.15i)]} + 0.5 {[B11 - (-0.20 + 0.00i)] + [B12 - (0.00 – 0.15i)] + [B21 - (0.00 + 0.00i)] + [B22 - (0.20 + 0.20i)]}. The second term is still set as Fbina = {1 - Σ($\varepsilon_i$ – $\varepsilon_1$) ($\varepsilon_i$– $\varepsilon_2$)}. Following the same optimization procedure as in the previous section, we have obtained the final patterns. Figure 4(a-b) demonstrates the final pattern of case 1 and case 2, which shows the pixel distribution within the device. We then conducted FDTD simulations for these patterns, which conclude the electric field in the device, the normalized power distributions of the output waveguides, and the normalized transmissions of TE0 and TE1 modes. As shown in Figure 4 (b), the TE0 mode is injected into the input waveguide (upper port or lower port) and mode conversions occur in the inverse-designed region, resulting in a superposition of TE0 and TE1 modes in the output waveguide. Specifically, the input fundamental modes are simultaneously converted into several orthogonal output modes. The simulation results indicate that the mode conversions between input modes and output modes are consistent with our design principle.

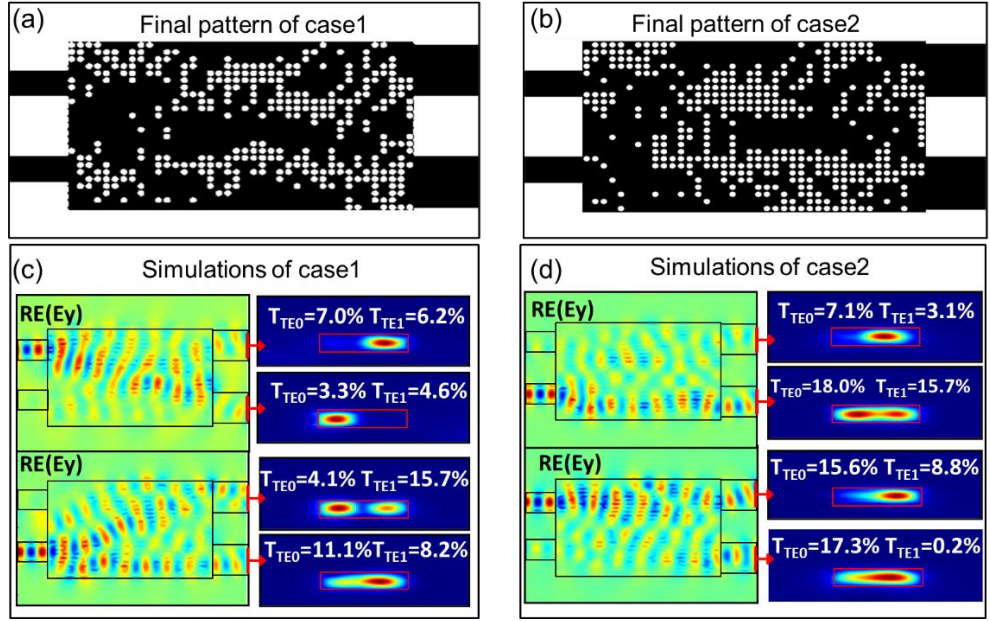

Figure 4: (a) Final pattern of the case 1. (b) Simulation results of the case 1, which conclude the electric fields, power distributions, and normalized transmissions. (c) Final pattern of the case 2. (d) Simulation results of the case 2, which conclude the electric fields, power distributions, and normalized transmissions.

To further evaluate the computing precisions of our device, we conducted FDTD simulations to obtain the design targets and the simulation values. We then draw a radar chart to compare the design target with the corresponding simulation values for the two matrices. Figure 4 (a-b) shows the radar chart for case 1 and case 2, respectively. Elements within the matrices are differentiated by colors, with targets represented by circles and simulations by crosses. The plot reveals that simulation values closely align with the design targets. Figure 4 (c-d) shows the statistical diagram and relative error analysis for case 1 and case 2, respectively. Specifically, the figure consists of three parts demonstrating the design target values, simulation values, and their relative error for the two conversion matrices. The simulation values exhibit slight deviations from the design target values for both conversion matrices. The relative errors for all the elements are no more than 8%, which represents the high computing precision of our photonics core. Through the analysis of these two cases, we can prove that our design principle is scalable for the parallel multiplication of two arbitrary matrices. This demonstrates that by setting the design targets, we can design a photonic core for the parallel multiplication of any specific matrix.

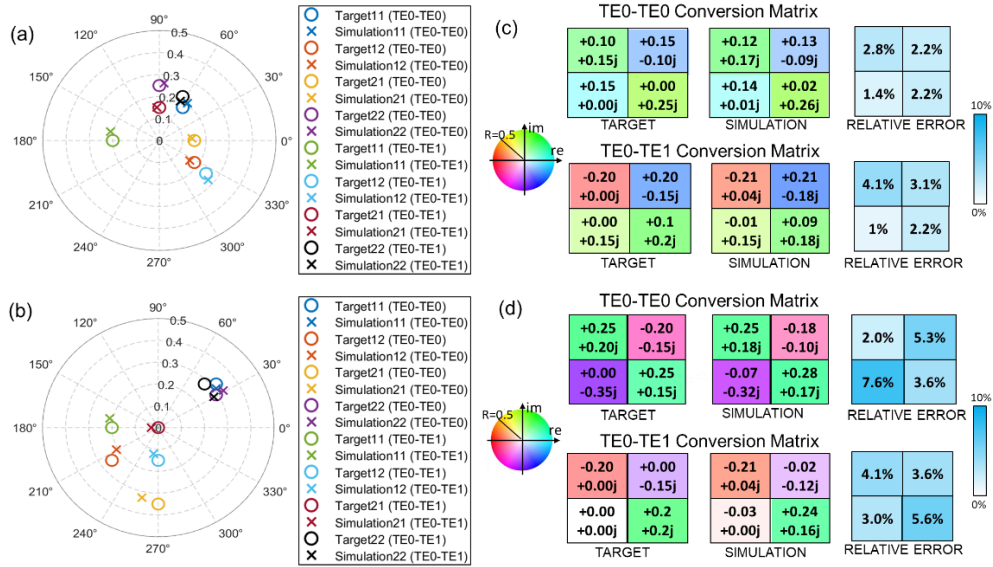

Figure 5: (a) Polar plot of case 1, which represents the design targets and simulated values for TE0-TE0 and TE0-TE1 mode conversion matrices. (b) Polar plot of case 2, which represents the design targets and simulated values for TE0-TE0 and TE0-TE1 mode conversion matrices. (c) Statistical diagram and relative error analysis of case 1 for TE0-TE0 conversion matrix and TE0-TE1 conversion matrix. (d) Statistical diagram and relative error analysis of case 2 for TE0-TE0 conversion matrix and TE0-TE1 conversion matrix.

## 5. Conclusion

In conclusion, we have proposed mode conversion-enabled photonic computing core for parallel matrix-vector multiplication using the inverse design method. The parallel matrix-vector multiplications are achieved through a mode conversion process, where the input fundamental modes are simultaneously converted into several orthogonal output modes. To fully utilize the orthogonality and the conversion characteristic of the waveguide modes, we employ the photonic-crystal-like (PhC-like) subwavelength structure and apply the adjoint method to inversely design the dielectric distribution, which shows powerful capacity in flexible control of waveguide mode. To demonstrate the scalability of our method, we have designed three device patterns: the first to realize parallel multiplication of zero matrix and diagonal matrices, and the subsequent two for the parallel multiplication of arbitrary matrices. The proposed photonic cores support the simultaneous parallel multiplication of two distinct independent matrices, with an ultra-compact footprint (2.4×2.88 µm²) and high computing precision (relative error < 8%) at the wavelength of 1550 nm. To our knowledge, this is the first photonic computing core that can exploit mode dimension for parallel matrix multiplication. We anticipate that by incorporating additional modes, our method could be extended to design parallel matrix-vector multiplication with higher degrees of parallelism and larger matrix dimensions.